\documentclass{article}

\usepackage{arxiv}

\usepackage[utf8]{inputenc} 
\usepackage[T1]{fontenc}    
\usepackage{hyperref}       
\usepackage{url}            
\usepackage{booktabs}       
\usepackage{amsfonts}       
\usepackage{nicefrac}       
\usepackage{microtype}      
\usepackage{lipsum}		
\usepackage{graphicx}
\usepackage{doi}
\usepackage{amsmath}
\usepackage{authblk}
\usepackage[title]{appendix}
\usepackage[sort&compress,numbers]{natbib}

\title{Solutions of the spray flamelet equations in a non-monotonic mixture fraction space}

\author[1]{Felipe Huenchuguala}
\author[1]{Luis Fuenzalida}
\author[2]{Oscar Orellana}
\author[3]{Arne Scholtissek}
\author[3]{\\Christian Hasse}
\author[4]{Eva Gutheil}
\author[ ~,1]{Hernan Olguin\thanks{Corresponding author\\ Email address: hernan.olguin@usm.cl}}

\affil[1]{Department of Mechanical Engineering, Universidad Técnica Federico Santa María,\linebreak Avenida España 1680, Valparaíso, Chile\linebreak}
\affil[2]{Department of Mathematics, Universidad Técnica Federico Santa María,\linebreak Avenida España 1680, Valparaíso, Chile\linebreak}
\affil[3]{Institute for Simulation of reactive Thermo-Fluid Systems, TU Darmstadt,\linebreak Otto-Berndt-Stra{\ss}e 2, 64287 Darmstadt, Germany\linebreak}
\affil[4]{Interdisciplinary Center for Scientific Computing, Heidelberg University,\linebreak Im Neuenheimer Feld 205, 69120 Heidelberg, Germany}

\date{}

\begin{document}
\maketitle

\begin{abstract}
Solving the spray flamelet equations in composition space is very challenging, which is attributable to the fact that the maximum value of the mixture fraction, $Z_\mathrm{max}$, is \textit{a priori} unknown in such flames. In this work, an analytical solution for this quantity is proposed, which allows its determination in spray flames subject to imposed quadratic evaporation profiles. It is then illustrated how the proposed approach allows to effectively cover the solution space of the spray flamelet equations. The employed strategy works very well for the considered cases and the generality of the evaporation profile definition provides flexibility for explorations of other parametric choices in the future.
\end{abstract}

\vspace{15mm}

\section{Introduction}

Different sets of flamelet equations are currently available in the literature for the description of spray flames in mixture fraction space~\cite{Luo13, Olguin14, Olguin143, Olguin142, Franzelli15, Olguin19, Maionchi21}. While their validity has been demonstrated by means of \textit{a priori} analyses of flames in physical space~\cite{Olguin14, Olguin143, Olguin142}, properly solving these equations in composition space can be very challenging. In particular, it is well-known that spray flames can present non-monotonic mixture fraction profiles with an \textit{a priori} unknown maximum value~\cite{Hollmann98,Watanabe07}, which also varies with the strain rate and the local evaporation characteristics. In other words, the domain in which the spray flamelet equations have to be solved is not \textit{a priori} known and, as a consequence, flamelet-based models for the simulation of turbulent spray flames still rely on counterflow flame structures solved in physical space~\cite{Olguin14,Hollmann98,Wang23,Hollmann96,Gutheil98,Ge081,Hu17,Hu172,Hu20,Wang232,Wang233}. However, this setup is not exempt of complexities and its adoption has introduced some additional issues. For example, it has been observed that, in this configuration, the droplet dynamics can lead to droplet reversal, which makes it difficult to systematically cover the entire solution space as required by flamelet tabulation methods. Additionally, it has been also shown that the counterflow flames can have multiple structures under identical boundary conditions~\cite{Gutheil05,Ying22,Ying23}.

In order to avoid the non-monotonicity problem, it has been proposed to solve the spray flamelet equations assuming a mixture fraction domain ranging from 0 to 1, which corresponds to a gas flamelet perturbed by a spray~\cite{Luo13, Olguin19}. However, this leads to reaction-diffusion dominated flamelet structures~\cite{Olguin19}, while previous works have demonstrated that evaporation induced convection is of major importance in spray combustion~\cite{Olguin14, Olguin143}, even surpassing diffusion in most cases. Alternatively, some works have proposed the use of modified definitions of the mixture fraction such as the effective composition variable~\cite{Franzelli15}, which combines the gaseous mixture fraction with the liquid-to-gas mass ratio, or the cumulative mixture fraction~\cite{Maionchi21}, defined as a weighted integral of the mixture fraction over the physical space domain. Although both variables are monotonic by definition, their implementation in flamelet-based models for the simulation of turbulent spray flames is not straightforward, which limits their applicability~\cite{Wang23}. Therefore, new strategies to address the spray flamelet equations in a systematic way are currently required.

The \textbf{main objective} of this work is introducing an approach to systematically solve the spray flamelet equations in a non-monotonic mixture fraction space, unlocking in this way their potential in the context of flamelet-based models for turbulent spray combustion. More specifically, an analytical solution for the mixture fraction is first obtained for flames subject to simplifications typically adopted in classical non-premixed flamelet theory (uniform strain rate, constant product of density and diffusion coefficients) and an imposed evaporation profile, $\dot{S}_v$, of the form $\dot{S}_v = Ay^2 + By + C$, where $A$, $B$ and $C$ are parameters. After demonstrating that the obtained solution reduces to the classical inverse error function profile when $\dot{S}_v = 0$, the spray flamelet equations are systematically solved for different strain rates and evaporation profiles, illustrating how the proposed approach can be employed in the generation of spray flamelet libraries. Finally, the budgets of the flamelet equations are analyzed to ensure that the previously observed reaction-diffusion-convection balance characterizing these flames is properly captured.


\section{Spray flamelet equations}\label{sec:flam_eq}

We start introducing the spray flamelet equations to be considered in this work. They are formulated in terms of the mixture fraction, $Z$, which is defined as a scalar satisfying the following transport equation~\cite{Olguin19}
\begin{equation}\label{eq:Z_phys_1}
\rho u\frac{dZ}{dy}=\frac{d}{dy}\left(\rho D_Z\frac{dZ}{dy}\right)+\dot{S}_v(1-Z),
\end{equation}
where $\rho$ denotes the gas density, $u$ the gas velocity, and $D_Z$ the diffusion coefficient of the mixture fraction, which is assumed to be equal to $\lambda/\rho C_p$ ($\mathrm{Le}_Z=1$), where $\lambda$ is the thermal conductivity and $C_p$ is the specific heat at constant pressure of the gas mixture. Based on $Z$, the equations for chemical species mass fractions, $Y_k$, and temperature, $T$, yield~\cite{Olguin19}
\begin{align}
0=
\underbrace{\frac{\rho}{\mathrm{Le}_k}\frac{\chi}{2}\frac{d^2Y_k}{d Z^2}}_\mathrm{I}
&+\underbrace{\dot{\omega}_k}_\mathrm{II}
+\underbrace{\dot{S}_v(\delta_{kF}-Y_k)}_\mathrm{III} -\underbrace{\dot{S}_v(1-Z)\frac{d Y_k}{d Z}}_\mathrm{IV}
+\underbrace{\Omega_{Y_k}}_\mathrm{V}
\label{eq:Yk}
\end{align} 
and
\begin{align}
0=
\rho C_p\frac{\chi}{2}\frac{d^2T}{d Z^2}
&+\dot{\omega}_T
+\dot{S}_e -C_p\dot{S}_v(1-Z)\frac{d T}{d Z}
+\Omega_T,\label{eq:T}
\end{align}
respectively. Here, $\chi$ is the scalar dissipation rate, and $\mathrm{Le}_k$ is the Lewis number of species $k$. Further, $\delta_{kF}$ represents the Kronecker delta, where the subscript $F$ refers to liquid fuel. The mass and energy sources due to evaporation are denoted by $\dot{S}_v$ and $\dot{S}_e$, while $\dot{\omega}_k$ and $\dot{\omega}_T$ are the chemical reaction rate and the energy source term due to chemical reactions, respectively. Finally, $\Omega_{Y_k}$ and $\Omega_T$ represent terms that, while important for the exact definition of the spray flamelet structures, are expected to be small. They are computed as
\begin{align}
\Omega_{Y_k}=&-\sqrt{\frac{\chi}{2D_Z}}\frac{d}{d Z}\left(\rho D_Z\left(\frac{\mathrm{Le}_k-1}{\mathrm{Le}_k}\right)\sqrt{\frac{\chi}{2D_Z}}\right)\frac{d Y_k}{d Z}  -\sqrt{\frac{\chi}{2D_Z}}\frac{d(\rho \widetilde{V}_{kZ}Y_k)}{d Z}
\label{eq:omegaYk}
\end{align}
and
\begin{align}
\Omega_T=\left(\rho \frac{\chi}{2}\frac{d C_p}{d Z}-\rho \sqrt{\frac{\chi}{2D_Z}}\sum_{k=1}^N C_{pk}V_{kZ}Y_k\right)\frac{d T}{d Z},
\end{align}
where $C_{pk}$ is the specific heat at constant pressure of species $k$, and $V_{kZ}$ and $\widetilde{V}_{kZ}$ are diffusion velocities in mixture fraction space
\begin{align}
V_{kZ} = -\sqrt{\frac{\chi}{2D_Z}}\frac{D_k}{Y_k}\frac{d Y_k}{d Z}+\widetilde{V}_{kZ}
\end{align}
and
\begin{align}
\widetilde{V}_{kZ} = -\sqrt{\frac{\chi}{2D_Z}}\frac{D_k}{\overline{W}}\frac{d \overline{W}}{d Z}-\sqrt{\frac{\chi}{2D_Z}}\frac{D_{Tk}}{\rho TY_k}\frac{d T}{d Z}+{V}_{kZ}^C.
\end{align}
In these equations, $D_k$ and $D_{Tk}$ denote the diffusion and thermal diffusion coefficients of species $k$ into the mixture, respectively, while $\overline{W}$ represents the mean molecular weight of the mixture. The velocity correction ensuring mass conservation, ${V}_{kZ}^C$, is
\begin{align}
V_{kZ}^C=\sqrt{\frac{\chi}{2D_Z}}\sum_{k=1}^N\left(\frac{D_kY_k}{X_k}\frac{d X_k}{d Z}+\frac{D_{Tk}}{\rho T}\frac{d T}{d Z}\right),
\end{align}
with $X_k$ denoting the mole fraction of species $k$.
 
For the closure of the scalar dissipation rate, $\chi=2D_Z\lvert dZ/dy \rvert^2$, a transport equation for the gradient of the mixture fraction, $g_Z=\lvert dZ/dy \rvert$, is considered~\cite{Olguin19}. In particular, the following equation will be adopted
\begin{align}
0=&g_Za+g_Z^2\frac{d}{d Z}\left(\frac{1}{\rho}\frac{d(\rho D_Zg_Z)}{d Z}\right)+g_Z^2\frac{d}{d Z}\left(\frac{\dot{S}_v(1-Z)}{\rho g_Z}\right). \label{eq:gz}
\end{align}
Given the assumption of a physical space coordinate $y$ coinciding with the direction of the gradient of the mixture fraction, the strain rate is defined in terms of the gas velocity in the direction normal to the mixture fraction iso-surfaces, $u$, as $a=-du/dy$~\cite{Olguin19}.

In order to solve Eqs.~\eqref{eq:Yk}, \eqref{eq:T} and \eqref{eq:gz}, boundary conditions need to be imposed at $Z=0$ and $Z=Z_\mathrm{max}$. However, as pointed out before, $Z_\mathrm{max}$ is an \textit{a priori} unknown quantity. In the next section, an analytical solution for $Z$ is presented, which provides a means to estimate its maximum value.


\section{Analytical solution for the mixture fraction}\label{sec:der}

In order to obtain an analytical solution for Eq.~\eqref{eq:Z_phys_1}, which then will be used to approximate the maximum value of the mixture fraction, $Z_\mathrm{max}$, it is assumed that: i) flamelets are established within a one-dimensional potential flow field with a local velocity $u=-ay$, and that ii) the product of $\rho$ and $D_Z$ is constant along the entire domain. With these considerations, which have also been employed in gas flames for the derivation of the classical analytical solutions of $\chi$~\cite{Peters84, Pitsch98}, the transport equation for $Z$ can be rewritten as
\begin{equation}\label{eq:Z_phys}
-\rho ay\frac{dZ}{dy}=\rho D_Z\frac{d^2Z}{dy^2}+\dot{S}_v(1-Z).
\end{equation}
Further, and as mentioned before, the evaporation source term will be assumed to have a general form characterized by a second degree polynomial as
\begin{equation}\label{eq:sv_pol}
\dot{S}_v(y)=Ay^2+By+C,
\end{equation}
where $A$, $B$ and $C$ are parameters.

Utilizing Eq.~\eqref{eq:sv_pol} together with a suitable mathematical transformation, Eq.~\eqref{eq:Z_phys} can be brought into a form having a known analytical solution, as will be shown next. Starting from Eq.~\eqref{eq:Z_phys}, we introduce a new dependent variable~\cite{Pala17}
\begin{equation}\label{eq:apSolPhiDefPsi}
\psi(y) = h(y)\exp\left(\int -\frac{1}{1-Z}\frac{dZ}{dy}dy\right),
\end{equation}
and a new coordinate
\begin{equation}\label{eq:nu}
\nu(y)=A^*y+\frac{2B}{{A^*}^3\rho D_Z},
\end{equation}
which leads to
\begin{equation}\label{eq:form}
\frac{d^2 \psi}{d\nu^2}-\left(\frac{1}{4}\nu^2+m\right)\psi=0,
\end{equation}
for which an analytical solution in terms of hypergeometric functions has been reported~\cite{Abramowitz65,Temme00}. Here, $h(y)$ is a function to be determined, and the coefficient $A^*$ is given by
\begin{equation}
A^*=\left(\frac{a^2}{D_Z^2}+\frac{4A}{\rho D_Z}\right)^{0.25}.
\end{equation} 

For the determination of $h(y)$, Eq.~\eqref{eq:apSolPhiDefPsi} is differentiated twice, which, after rearranging terms, leads to
\begin{align}\label{eq:apSolPhiRicattiMN}
\frac{d^2Z}{dy^2} = - \frac{2h'}{h}\frac{dZ}{dy} -\frac{(1-Z)}{\psi}\frac{d^2\psi}{dy^2} + (1-Z)\frac{h''}{h},
\end{align}
where $h'$ and $h''$ denote the first and second derivative of $h$, respectively. Additionally, a second expression for $d^2Z/dy^2$ can be obtained from Eq.~\eqref{eq:Z_phys}, which yields
\begin{equation}\label{eq:apSolPhiRicatti}
\frac{d^2Z}{dy^2}=-\frac{ay}{D_Z}\frac{dZ}{dy}-\frac{\dot{S}_v(1-Z)}{\rho D_Z}.
\end{equation}
Comparing the coefficients of the terms involving the first derivative of $Z$ in Eq.~\eqref{eq:apSolPhiRicattiMN} and Eq.~\eqref{eq:apSolPhiRicatti}, the function $h(y)$ is specified as
\begin{align}
h(y) &= \exp\left(\frac{ay^2}{4D_Z}\right). \label{eq:MSolution}
\end{align}

Further, equating the last two terms on the RHS of Eq.~\eqref{eq:apSolPhiRicattiMN} with the last term on the RHS of Eq.~\eqref{eq:apSolPhiRicatti} (and inserting Eq.~\eqref{eq:sv_pol}), the following transport equation for $\psi$ is obtained
\begin{align}\label{eq:apSolPhiPsiEquation}
\frac{d^2\psi}{dy^2} - \left[ \left(\frac{a^2}{4D_Z^2} + \frac{A}{\rho D_Z}\right) \right. &y^2 + \frac{B}{\rho D_Z} y \left. +\frac{a}{2D_Z} + \frac{C}{\rho D_Z} \right]\psi = 0,
\end{align}
where
\begin{align}\label{eq:psi_y}
\psi(y)=\exp\left[\frac{ay^2}{4D_Z} -\int \frac{1}{1-Z}\frac{dZ}{dy} dy\right].
\end{align}

Now we introduce a change of coordinate from $y$ to $\nu$, which is formalized through
\begin{equation}\label{eq:y(psi)}
y=\frac{\nu}{A^*}-\frac{2B}{{A^*}^4\rho D_Z}
\end{equation}
and
\begin{equation}\label{eq:ddy}
\frac{d^2}{dy^2}={A^*}^2\frac{d^2}{d\nu^2}.
\end{equation}
Substituting Eqs.~\eqref{eq:y(psi)} and~\eqref{eq:ddy} into Eq.~\eqref{eq:apSolPhiPsiEquation}, we obtain the desired form for the $\psi$-equation, which yields
\begin{equation}\label{eq:psi_nu}
\frac{d^2\psi}{d\nu^2}-\left(\frac{1}{4}\nu^2+m\right)\psi=0,
\end{equation}
where the parameter $m$ is given by
\begin{equation}
m=\frac{1}{{A^*}^2}\left(\frac{a}{2D_Z}+\frac{C}{\rho D_Z}-\left(\frac{B}{{A^*}^2\rho D}\right)^2\right).
\end{equation}
As pointed out before, an analytical solution for this equation already exists. In particular, the corresponding linearly independent solutions of Eq.~\eqref{eq:psi_nu} can be written as~\cite{Abramowitz65, Temme00}
\begin{align}
\psi_1(y)&=\exp\left(-\frac{\nu(y)^2}{4}\right) {}_1F_1 \left(\frac{1}{4}+
\frac{m}{2},\frac{1}{2},\frac{\nu(y)^2}{2}\right)
\end{align}
and
\begin{align}
\psi_2(y)&=\nu(y)\exp\left(-\frac{\nu(y)^2}{4}\right) {}_1F_1 \left(\frac{3}{4}+
\frac{m}{2},\frac{3}{2},\frac{\nu(y)^2}{2}\right),
\end{align}
where ${}_1F_1$ is the cofluent hypergeometric function of the first kind, defined as
\begin{equation}\label{eq:cofluent}
{}_1F_1(a,b,x)=\sum_{n=0}^\infty\frac{(a)_n}{(b)_n}\frac{x^n}{n!},
\end{equation}
where $(a)_n$ is the Pochhammer symbol (rising factorial), given by
\begin{equation}
(a)_n=\frac{\Gamma(a+n)}{\Gamma(a)},
\end{equation}
and $\Gamma$ is the gamma function.

At this point, Eq.~\eqref{eq:psi_y} can be inverted to obtain an explicit analytical expression for $Z$. In particular, it is easy to show that
\begin{equation}
\frac{1}{1-Z}\frac{dZ}{dy}=\frac{ay}{2D_Z}-\frac{1}{\psi(y)},
\end{equation}
and, after separation of variables and integration, $Z$ can be determined as a function of $y$ as
\begin{equation}
Z(y)=1-\exp\left(-\frac{ay^2}{4D_Z}\right)\psi(y).
\end{equation}
Finally, expressing $\psi(y)$ as a linear combination of the two independent solutions $\psi_1(y)$ and $\psi_2(y)$, we obtain
\begin{equation}\label{eq:Z_y}
Z(y)=1+\exp\left(\frac{-ay^2}{4D_Z}\right) \left(\alpha\psi_1(y)+\beta\psi_2(y)\right),
\end{equation}
where $\alpha$ and $\beta$ are constants to be determined from the specific boundary conditions. 


\section{Consistency with classical solutions for gas flames}
\label{sec:consistency}

Before proceeding to a demonstration of how the analytical solution developed in the previous section can be used to solve the spray flamelet equations in a non-monotonic mixture fraction space, it will be shown that Eq.~\eqref{eq:Z_y} reduces to the classical erfc-profile for $\chi$ when $\dot{S}_v(y)=0$ ($A$, $B$ and $C$ are set to zero, cf. Eq.~\eqref{eq:sv_pol}). For that case, $m=1/2$ and $\nu(y)=\sqrt{a/D_Z}y$, which allows expressing $\psi_1(y)$ and $\psi_2(y)$ as
\begin{equation}
\psi_1(y)=\exp\left(-\frac{ay^2}{4D_Z}\right) {}_1F_1 \left(\frac{1}{2},\frac{1}{2},\frac{ay^2}{2D_Z}\right) \label{eq:phi1gas}
\end{equation}
and
\begin{equation}
\psi_2(y)=\sqrt{\frac{a}{D_Z}}y\exp\left(-\frac{ay^2}{4D_Z}\right) {}_1F_1 \left(1,\frac{3}{2},\frac{ay^2}{2D_Z}\right) ,\label{eq:phi2gas}
\end{equation}
respectively, with~\cite{Abramowitz65}
\begin{equation}\label{eq:1f11}
\vphantom{F}_1F_1\left(\frac{1}{2},\frac{1}{2},\frac{ay^2}{2D_Z}\right) = \exp\left(\frac{ay^2}{2D_Z}\right)
\end{equation}
and
\begin{align}\label{eq:1f12}
\vphantom{F}_1F_1\left(1,\frac{3}{2},\frac{ay^2}{2D_Z}\right)& = 
\sqrt{\frac{\pi D_Z}{2a}}\frac{1}{y}\exp\left(\frac{ay^2}{2D_Z}\right)\mathrm{erf}\left(\sqrt{\frac{a}{2D_Z}}y\right).
\end{align}

Inserting expressions~\eqref{eq:1f11} and~\eqref{eq:1f12} into Eq.~(\ref{eq:Z_y}), and applying the following boundary conditions~\cite{Peters84}
\begin{align}
Z(y\rightarrow -\infty)&=1\\
Z(y\rightarrow \infty)&=0,
\end{align}
leads to
\begin{equation}\label{eq:xigas}
Z(y) = \frac{1}{2}-\frac{1}{2}\mathrm{erf}\left(\sqrt{\frac{a}{2D_Z}}y\right) = \frac{1}{2}\mathrm{erfc}\left(\sqrt{\frac{a}{2D_Z}}y\right), 
\end{equation}
which implies
\begin{equation}\label{eq:ggas1}
g_{Z}(y) = \sqrt{\frac{a}{2D_Z\pi}}\exp\left(-\frac{ay^2}{2D_Z}\right).
\end{equation}
After inverting $Z(y)$, $g_Z$ can be expressed as a function of mixture fraction as
\begin{equation}
g_Z(Z) = \sqrt{\frac{a}{2D_Z\pi}}\exp\left(-\left(\mathrm{erfc}^{-1}(2Z)\right)^2\right),
\end{equation} 
which can be used to obtain
\begin{equation}
\chi = \frac{a}{\pi} \exp \left(- 2\left(\mathrm{erfc}^{-1}(2Z)\right)^2\right),
\end{equation}
the classical expression for the scalar dissipation rate used in the context of non-premixed gas flamelets~\cite{Peters84}.


\section{Proof of concept}\label{sec:strc}

Now it will be illustrated how the proposed analytical expression for $Z$, Eq.~\eqref{eq:Z_y}, can be used to solve the spray flamelet equations in a non-monotonic composition space. For simplicity, a symmetry condition will be assumed at $Z_\mathrm{max}$, which can be interpreted as a twin counterflow spray flame. It is remarked, however, that the formulation is general and it can also be applied to asymmetric flames, with a spray being injected from only one side of the configuration.

Formally, boundary conditions of Dirichlet type are imposed at $Z=0$ for Eqs.~\eqref{eq:Yk}, \eqref{eq:T}, and \eqref{eq:gz}, corresponding to the injection of pure air at $300$~K and atmospheric pressure. These are given by
\begin{equation}\label{eq:cond_1}
Z=0:\quad Y_k=Y_{k,0}, \quad T=300\,\mathrm{K}, \quad g_Z=0.
\end{equation}
At $Z=Z_\mathrm{max}$, on the other hand, boundary conditions are expressed as
\begin{equation}\label{eq:cond_2}
Z=Z_\mathrm{max}:\quad \frac{dY_k}{dZ}=0, \quad \frac{dT}{dZ}=0, \quad g_Z=0.
\end{equation}
To determine $Z_\mathrm{max}$, a physical domain ranging between $-\infty$ and $0$ is considered. Given the previously assumed velocity profile ($u=-ay$), $y=0$ corresponds to the stagnation point. Further, with Eqs.~\eqref{eq:cond_1} and~\eqref{eq:cond_2}, the boundary conditions needed to determine $\alpha$ and $\beta$ in Eq.~\eqref{eq:Z_y} become 
\begin{align}
Z(y\rightarrow-\infty) &= 0
\end{align}
and
\begin{align}
\left.\frac{dZ}{dy} \right|_{y=0} &= 0,\label{eq:cond_gz}
\end{align}
respectively.

In order to specify the evaporation mass source term, the constant coefficients $A$, $B$ and $C$, appearing in Eq.~\eqref{eq:sv_pol}, are prescribed as
\begin{equation}\label{eq:sv_const}
A= -\frac{4 \dot{S}_{v, \mathrm{max}}}{l_f^2}, \quad B=-\frac{4 \dot{S}_{v, \mathrm{max}}}{l_f}, \quad C=0,
\end{equation}
where, $l_f$ is the flamelet length defined as the region of the domain in which the mixture fraction has a value different from zero. With Eq.~\eqref{eq:sv_const}, $\dot{S}_v$ and $\dot{S}_e$ can be expressed as
\begin{equation}\label{eq:sv}
\dot{S}_v(y) = -\frac{4 \dot{S}_{v, \mathrm{max}}}{l_f^2} y^2
-\frac{4 \dot{S}_{v, \mathrm{max}}}{l_f} y ,
\end{equation}
and
\begin{equation}\label{eq:se}
\dot{S}_e(y) = -\dot{S}_v(y)\left(C_{pF}(T-T_\mathrm{boil})+L_v\right) ,
\end{equation}
respectively, where $T_\mathrm{boil}$ denotes the boiling temperature of the liquid fuel, and $L_v$ the latent heat of vaporization. In the remainder of this section, the maximum evaporation rate, $\dot{S}_{v,\mathrm{max}}$, will be treated as a free parameter allowing to systematically generate the different flamelet structures of interest.

The spray flamelet equations are solved now for three different strain rates, $200$/s, $300$/s and $400$/s, while $\dot{S}_{v,\mathrm{max}}$ is varied from $30$ to $80$~kg/(m$^3$s). The algorithm employed for this includes the following steps:
\begin{itemize}
\item[S$_1$:] For a given combination of $a$ and $\dot{S}_{v,\mathrm{max}}$, a first estimation of the $Z_\mathrm{max}$ value is obtained from Eq.~\eqref{eq:Z_y}.
\item[S$_2$:] Initial guesses for the $Y_k$, $T$ and $g_Z$ profiles are assumed based on previous computations. The latter is then used to estimate the internal distribution of the coordinate $y$ within the flame, as well as $l_f$ (see Eq.~\eqref{eq:sv}). With these values, the evaporation profiles are calculated and projected into composition space.
\item[S$_3$:] The flamelet equations, Eqs.~\eqref{eq:Yk},~\eqref{eq:T} and~\eqref{eq:gz}, are solved, and the new $g_Z$-profile is employed to recalculate $l_f$ to modify the $\dot{S}_v$ and $\dot{S}_e$ profiles according to Eqs.~\eqref{eq:sv} and~\eqref{eq:se}, to recalculate $Z_\mathrm{max}$, and to correct their corresponding projections into mixture fraction space. This step is repeated until the relative error of all the flamelet equations and $Z_\mathrm{max}$ between iterations are lower than a convergence value, $\epsilon$.
\end{itemize}

\begin{figure*}[t!]
\centering
\includegraphics[width=.97\textwidth]{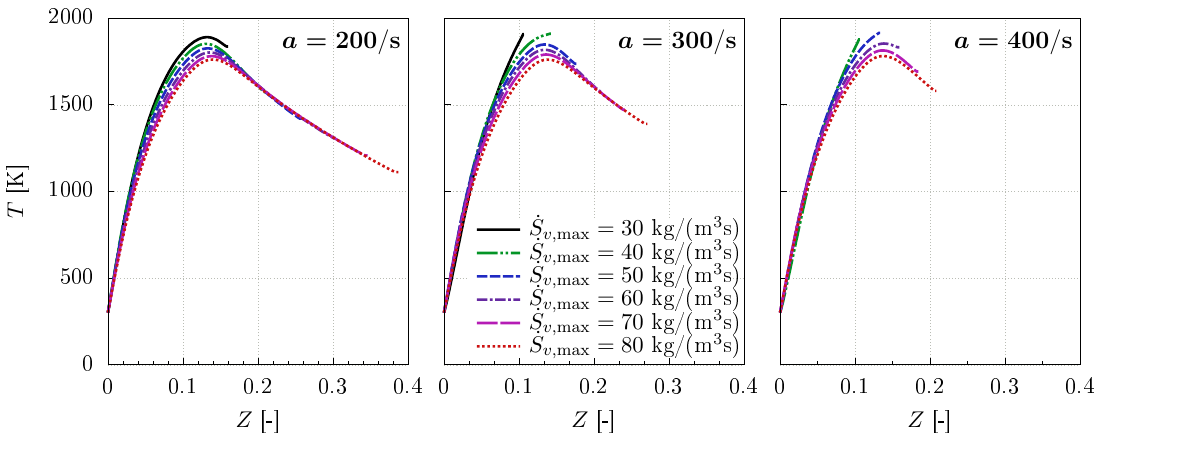}
\caption{Temperature profiles of flamelets corresponding to different values of $\dot{S}_{v,\mathrm{max}}$, at strain rates of $a=200$/s (left), $300$/s (center), and $400$/s (right)}
\label{fig1}
\end{figure*}
\begin{figure*}[t!]
\centering
\includegraphics[width=.97\textwidth]{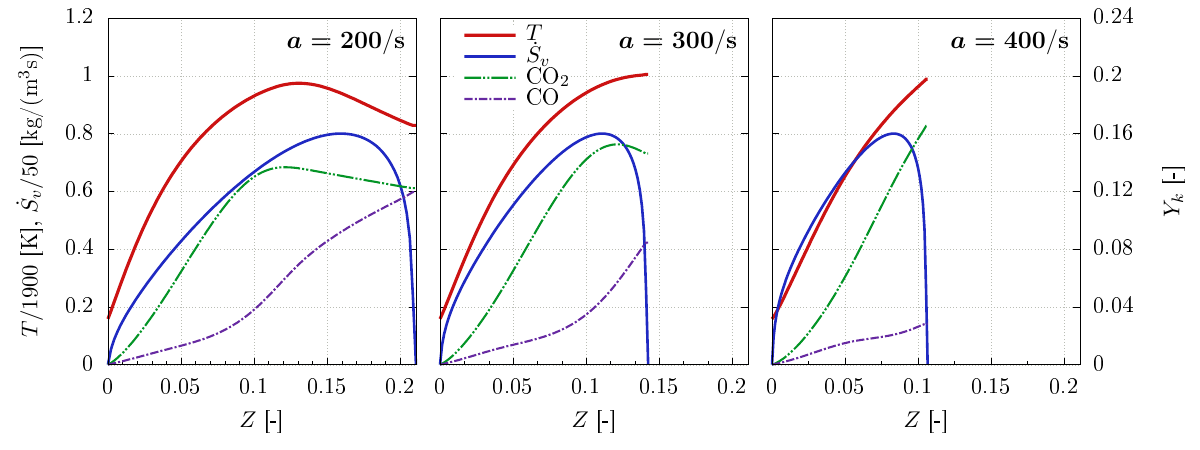}
\caption{Profiles of $T$, $\dot{S}_v$, $Y_\mathrm{CO}$, and $Y_{\mathrm{CO}_2}$ for flamelets with $\dot{S}_{v,\mathrm{max}}=40$~kg/(m$^3$s), at strain rates of $a=200$/s (left), $300$/s (center), and $400$/s (right).}
\label{fig2}
\end{figure*}
\begin{figure*}[t!]
\centering
\includegraphics[width=.97\textwidth]{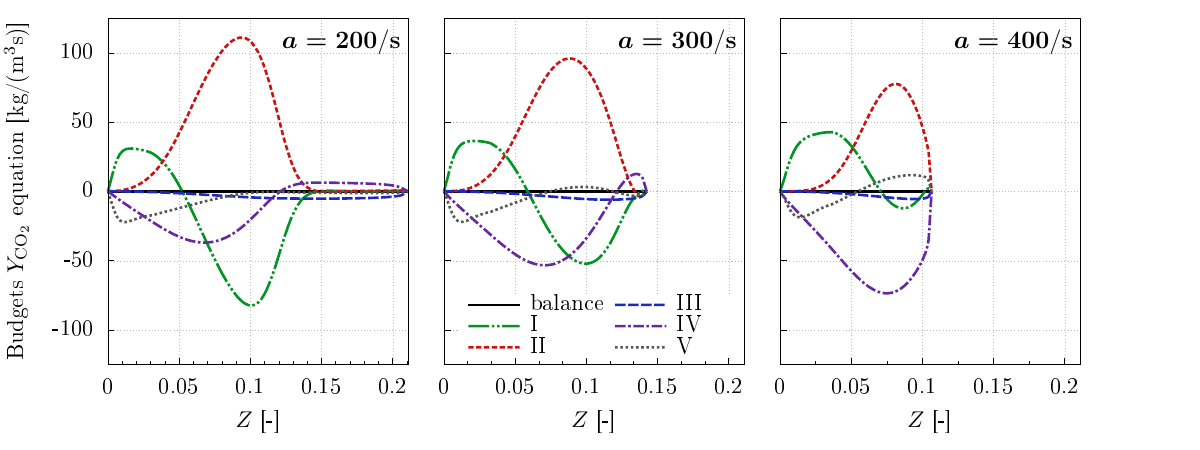}
\caption{Budgets of the $Y_{\mathrm{CO}_2}$ equation, Eq.~\eqref{eq:Yk}, for flamelets with $\dot{S}_{v,\mathrm{max}}=40$~kg/(m$^3$s) at strain rates of $a=200$/s (left), $300$/s (center), and $400$/s (right).}
\label{fig3}
\end{figure*}

As it can be seen in Fig.~\ref{fig1}, the adopted procedure allows systematically covering the solution space of the spray flamelets equations. Further, as shown in Fig.~\ref{fig2} for three selected flamelets, the obtained structures exhibit the typical $T$, $Y_{\mathrm{CO}_2}$ and $Y_\mathrm{CO}$ distributions previously observed in the context of counterflow spray flames~\cite{Olguin14,Olguin143}. In particular, a transition is observed from two temperature peaks at low strain rate (one at each side of the symmetry axis) to a single peaked profile at higher values of $a$ (located at the point of maximum mixture fraction). Additionally, the corresponding budgets of the spray flamelet equations, Fig.~\ref{fig3}, show the expected reaction-diffusion-convection competition in which the latter effect considerably gains in relevance when strain rate is increased. As pointed out above, previous attempts of solving the spray flamelet equations in a mixture fraction space ranging between zero and unity (gas flamelets perturbed by a spray) could not properly capture this behavior, predicting reaction-diffusion dominated flamelet structures.

As a final note, it is highlighted that the evaporation profiles employed here are not the only possible choices and that several different alternatives exist for they prescription. In this sense, the current work opens the door for future studies aiming to determine the best strategies for the generation of flamelets libraries for the simulation of turbulent spray flames.


\section{Conclusions}\label{sec:con}

In this work, the spray flamelet equations have been solved in a non-monotonic mixture fraction space for the first time. In order to enable this, an analytical solution for $Z$ was derived for imposed evaporation profiles of the form $\dot{S}_v = Ay^2 + By + C$, which allowed calculating $Z_\mathrm{max}$ and properly imposing the required boundary conditions in this domain. A proof of concept was then carried out, illustrating how the proposed approach allows systematically covering the solution space of the spray flamelet equations by varying the gas strain rate and the peak of $\dot{S}_v$. It was also demonstrated that the generated flamelet structures exhibited the main features of counterflow spray flames previously reported in the literature and that their budgets describe the expected reaction-diffusion-convection internal nature that could not be capture in previous works by analyzing gas flamelets perturbed by a spray in the entire mixture fraction domain ranging between zero and unity.

The present approach has a great potential of application in the elaboration of laminar spray flame libraries for the simulation of turbulent spray flames and, as future work, different strategies for the systematic prescription of evaporation profiles needs to be studied.


\section*{Acknowledgments}

The authors acknowledge funding from Proyecto Interno USM PI\_LIR\_2022\_15, FH thanks ANID (Chile) for financial support through the Doctorate Scholarship 21201826.

\bibliography{references}

\end{document}